# Protection gaps in Amazon floodplains will increase with climate change: Insight from the world's largest scaled freshwater fish

Nicolas Dubos[1*], Maxime Lenormand[1], Leandro Castello[2], Thierry Oberdorff[3], Antoine Guisan[4], Sandra Luque[1]

[1] TETIS, Univ Montpellier, AgroParisTech, Cirad, CNRS, INRAE, Montpellier, France

[2] Department of Fish and Wildlife Conservation, Virginia Polytechnic Institute and State University, Blacksburg, VA, USA

[3] UMR EDB (Laboratoire Évolution et Diversité Biologique), CNRS 5174, IRD 253, UPS, 118 route de Narbonne, F-31062 Toulouse, France

[4] Department of Ecology and Evolution, University of Lausanne, Lausanne, Switzerland

* Corresponding author: Nicolas Dubos, dubos.research@gmail.com

## Abstract

1. The Amazon floodplains represent important surfaces of highly valuable ecosystems, yet they remain neglected from protected areas. Although the efficiency of the protected area network of the Amazon basin may be jeopardized by climate change, floodplains are exposed to important consequences of climate change but are omitted from species distribution models and protection gap analyses.

2. The present and future (2070) distribution of the giant bony-tongue fish *Arapaima* spp. (Arapaimidae) was modelled accounting for climate and habitat requirements, and with a consideration of dam presence (already existing and planned constructions) and hydroperiod (high- and low-water stages). The amount of suitable environment that falls inside and outside the current network of protected areas was quantified to identify spatial conservation gaps.

3. We predict that climate change will cause a decline in environmental suitability by 16.6% during the high-water stage, and by 19.4% during the low-water stage. About 70% of the suitable environments of *Arapaima* spp. remain currently unprotected. The gap is higher by 0.7% during the low-water stage. The lack of protection is likely to increase by 5% with future climate change effects. Both existing and projected dam constructions may hamper population flows between the central, Bolivian and Peruvian parts of the basin.

4. We highlight protection gaps mostly in the south-western part of the basin and recommend the extension of the current network of protected areas in the floodplains of the upper Ucayali, Juruà and Purus rivers and their tributaries. This study has shown the importance of integrating hydroperiod and dispersal barriers in forecasting the distribution of freshwater fish species, and stresses the urgent need to integrate floodplains within the protected area networks.

**Keywords:** *Arapaima* spp., Climate change, Dams, Gap analysis, Hydroperiod, Species Distribution Models, Water colour.

# Introduction

The Amazon basin is the world's largest basin and hosts the highest freshwater fish species richness on Earth (Oberdorff et al., 2019; Jézéquel et al., 2020a). Its fish fauna is currently under multiple threats such as habitat fragmentation by dams, deforestation, urban and agricultural pollutants, and overexploitation (Reid et al., 2019; Duponchelle et al., 2021). Climate change may exacerbate these threats by increasing water temperatures and changing precipitation patterns, leading to increasing precipitations in the western part of the basin and to more severe and longer droughts in its southern and southeastern portion (Nobre et al., 2016; Sorribas et al., 2016; Marengo et al., 2018) potentially affecting the Amazon's fish fauna (Tedesco et al., 2013).

During The International Union for Conservation of Nature World Congress (Marseille, France, 5 September 2021), indigenous groups urged the world leaders to take action to protect 80% of the Amazon basin by 2025, stating that bold action is needed to halt the deforestation that is pushing the world's largest rainforest beyond the point of no return. The present network of protected areas (Pas, including indigenous lands) of the Amazon Basin was mainly designed for terrestrial species, which has downplayed the importance of aquatic ecosystems (Dagosta et al., 2021). Protected areas and indigenous lands cover 52% of the basin surface (Jézéquel et al., 2020b), which is potentially sufficient to encompass the present distribution of a majority of large-range fish species (Frederico et al., 2021). However, the capacity of the network to protect freshwater biodiversity remains limited (Azevedo-Santos et al., 2019; Jézéquel et al., 2020b). In addition, following future climate change projections, the proportion of species with a sufficient area protected will decrease significantly by 2050 as a result of species distribution shifts (Frederico et al., 2021). The greatest protection gaps are located in large rivers and their floodplains, emphasizing the need to focus conservation efforts in these habitats and identify areas where species are predicted to shift (Frederico et al., 2021). Fish species represent a substantial part of the Amazonian diversity (>2200 species described; Jézéquel et al., 2020a) and play major roles in Amazonian ecosystems. The omission of aquatic areas, especially floodplains, from PA networks may increase the basin's vulnerability to ongoing changes.

Species of the giant bony-tongue fish *Arapaima* spp., also known as *pirarucú* in Brazil or *paiche* in Peru and Bolivia, form an obligate air-breathing omnivorous species complex with piscivorous tendencies (Carvalho et al., 2018). *Arapaima* spp. represents the world's largest scaled freshwater taxon and one of the most charismatic fish of the Amazon. Here *Arapaima* spp. is considered a species complex, as the genus may not be monospecific, being potentially constituted by two or three closely related species (Stewart, 2013a; Stewart 2013b; but see Torati et al., 2019). *Arapaima* spp. migrate laterally for feeding and reproduction between river channels and floodplain habitats following seasonal hydrodynamics (Castello, 2008). Lateral migrant species such as *Arapaima* spp. are considered vulnerable to climate change because floodplain lakes are highly exposed to temperature changes (Duponchelle et al., 2021). *Arapaima* spp. is therefore a good flagship model to better understand the effects of climate change on lowland Amazonian fish species displaying lateral migrations. It is also a key fishing resource eliciting a high economic value in the region (Castello et al., 2009; Macnaughton et al., 2015). Hence, protecting the *Arapaima* spp. range should benefit to an important component of biodiversity (e.g. floodplain-dwelling species) while securing food resource for local communities (Petersen et al., 2016).

Conservation recommendations for management usually account for the species habitat requirements (Arantes et al., 2010; Arantes et al., 2013). However, conservation areas may be appropriate under present conditions, but not in the future (Leroy et al., 2014;

Frederico et al., 2021). Therefore, there is a need to assess the representativeness of PAs for *Arapaima* spp. in regard to potential distribution shifts driven by future climate change. To date, only two studies have examined the potential effects of future climate change on *Arapaima* spp. distribution (Oberdorff et al., 2015; Oliveira et al., 2020). The first one used a reduced set of species occurrence points and only one climatic variable (Oberdorff et al. 2015), while the other was mainly focused on historical demography and genomic diversity (Oliveira et al*.*, 2020). Hence, there is a need to develop forecasting models that are dedicated to the design of conservation actions for lowland floodplain-dwelling species in general and *Arapaima* spp. in particular.

Climate change effects are commonly predicted using species distribution models (also called ecological niche or habitat suitability models; Guisan, Thuiller & Zimmermann, 2017). This approach has often proven effective in conservation planning (e.g. providing recommendations for translocation, habitat restoration or the design of PAs; Leroy et al., 2014; Dubos et al., 2022b; Dubos et al., 2022a). Reliable species distribution models are demanding in their methodological design and data input (Guisan et al., 2013; Araújo et al., 2019; Sofaer et al., 2019; Sillero et al., 2021). Ideal conditions for reliable modelling may hardly be reached, yet species distribution model were founnd to be effective in the context of urgent decision-making (Guisan et al. 2013). The aim of this article is to provide guidelines for the design of present and future (2070) PAs on the basis of acceptable methodological considerations for decision-making, using the most recently available data and ecologically relevant environmental variables, treating sample bias, considering hydroperiods and dispersal barriers, and dealing with model complexity.

## Methods

*Study species and area*

*Arapaima* spp. is naturally distributed in the sub-basins of the Amazon, Tocantins-Araguaia and Essequibo rivers, which cover Brazil, Ecuador, Colombia, Guyana and Peru (Castello & Stewart, 2010). Within the Amazon River basin and according to available data (Jézéquel et al., 2020a), *Arapaima* spp. are naturally evenly distributed in highly productive Amazonian floodplain nutrient-rich white (turbid) waters (Fernandes, Podos & Lundberg, 2004), with the notable exception of the upstream section of the Madeira River (Bolivian Amazon), where a series of rapids probably historically acted as barriers to colonization (Miranda-Chumacero et al., 2012). However, *Arapaima* spp. colonized Bolivian waters (where it is now considered an invasive species) following an unintentional introduction around the late 1970s via the Peruvian side of the Madre de Dios River (Miranda-Chumacero et al., 2012).

*Occurrence data*

In total, 172 occurrence records available from a recent fish occurrence database were retrieved, gathering all information available in published articles, books, gray literature, online databases, foreign and national museums, and universities (Jézéquel et al., 2020a). To reduce spatial autocorrelation, one occurrence per pixel of the environmental variables (30 arcsec) was selected, resulting in 162 presence points (i.e. data thinning; Vollering et al., 2019; Steen et al., 2021).

*Climate data*

The 11 bioclimatic variables of temperature at 30 arcsec resolution (approximately 900 m) of the present climate data and the 2070 projections from CHELSA (Karger et al., 2017) were used. Only temperature variables were taken because freshwater fishes are particularly sensitive to temperature via direct effects on their metabolism, development and reproduction (Buisson, Blanc & Grenouillet, 2008), and because precipitation data tend to be less accurate in riverine systems, especially in regions with heavy rainfall and a low density of meteorological stations (Soria-Auza et al., 2010; Woldemeskel et al., 2015). This is particularly true for future projections (Kent, Chadwick & Rowell, 2015; Kim et al., 2020; Marra et al., 2021). Three Global Circulation Models (GCMs – i.e., BCC-CSM1-1, MIROC5 and HadGEM2-AO) and two greenhouse gas emission scenarios (shared socio-economic Pathways; also called representative concentration pathways, RCPs) were used, the most optimistic RCP2.6 and the most pessimistic RCP8.5. In the case of aquatic species, climate variables can be used as satisfactory surrogates for unavailable instream variables (e.g. water temperature) to model their distributions because water and air temperatures are usually highly correlated (Frederico, De Marco & Zuanon, 2014; McGarvey et al., 2018).

*Distribution modelling*

The distribution of *Arapaima* spp. was modelled with the Biomod2 R package (Thuiller et al., 2009), using 10 modelling techniques: generalized linear and generalized additive models (Guisan, Edwards & Hastie, 2002); classification tree analysis (Prasad, Iverson & Liaw, 2006); artificial neural network (Manel, Dias & Ormerod, 1999); surface range envelop (also known as BIOCLIM; Booth et al., 2014); flexible discriminant analysis (Manel, Dias & Ormerod, 1999); random forest (Prasad, Iverson & Liaw, 2006); multiple adaptive regression splines (Leathwick et al., 2005); generalized boosting model (Elith, Leathwick & Hastie, 2008); and maximum entropy (Phillips, Dudik & Schapire, 2006). Five different sets of 10,000 randomly selected pseudo-absences (Wisz & Guisan, 2009) down-weighted to equal presence data (setting prevalence to 0.5; Liu, Newell & White, 2019) were generated. One variable per group of inter-correlated variables was selected to avoid collinearity (Pearson's r > 0.7, Dormann et al., 2013) using the removeCollinearity function of the virtualspecies R package (Leroy et al., 2016). The relative importance of each variable kept was assessed with 10 permutations per modelling technique and pseudo-absence sets (see below; 10 × 10 × 5 = 500 total permutations; for details see BIOMOD package documentation available at http://r-forge.r-project.org/projects/biomod/). The variables included in the final models were those with a relative importance > 0.2 across at least 50% of model runs. Species distributions were predicted with an ensemble of small models approach (Lomba et al., 2010; Breiner et al., 2015). Sets of bivariate models were run, i.e. including all pairwise combinations of the selected variables, and an ensemble model with the mean predictions across all models weighted by their respective Boyce index (see below) was produced. This method is advocated for data-poor species and enables to reduce model complexity without reducing the explanatory power. The Amazon basin was taken as a background, defined as the area of land where precipitation collects and drains off into a common outlet (i.e. the Atlantic Ocean). This excludes de facto the Tocantins basin and Guiana coastal streams (Mayorga et al., 2005; Jézéquel et al., 2020a) but allows to work on a closed, ecologically meaningful system (Dudgeon, 2019). Ideally, the background extent should represent the area within which the species is able to disperse, i.e. rivers and wetlands. However, the whole Amazon Basin was chosen as a background and not only the rivers and wetlands because this would require either: (i) to downscale the climatic data at the resolution of wetland variables, inducing interpolations which are not recommended in most cases (Sillero & Barbosa, 2021); or (ii) to aggregate wetland data, which would represent a significant loss of information (i.e. most tributaries would be removed because they cover the minority of the pixels' surface). Therefore, aquatic habitat was accounted for by post-filtering the projections to remove strictly terrestrial areas (see *Habitat and land use data* section below).

Most occurrence data were obtained from dedicated local studies and may be subject to sample bias. To account for potential sample biases, five additional sets of pseudo-absences were generated around the original (unthinned) presence points (Phillips et al., 2009). A geographic null model (i.e. a map of the geographic distance to presence points; Hijmans, 2012) was used as a probability weight for pseudo-absence generation. The data for performance evaluation was partitioned spatially, using 5-folds for block-cross validation (generated with the blockCV R package, Valavi et al., 2019; Figure S1). The effect of sample bias correction (non-random pseudo-absence generation) was quantified using the relative overlap index (Dubos et al., 2021). This index informs how sample bias corrections affected spatial predictions (mean Schoener's D between uncorrected and corrected individual models) relative to inter-model variability (mean overlap between all pairwise combinations of model replicates). A value of 1 indicates a strong effect of correction across all model replicates (pseudo-absence and block cross-validation runs), whereas a value below 0 indicates that correction effect is lower than the variability between model replicates. This index was computed independently for each modelling technique and ensemble of small models and gives the mean value.

Model performance was assessed using three widely used evaluation metrics (for comparability), i.e. the area under the curve of a receiver operating characteristic plot (AUC; Swets & Swets, 1988), a maximisation of the true skill statistics (maxTSS; Allouche, Tsoar & Kadmon, 2006) and the Boyce index (Hirzel et al., 2006). AUC and TSS are discrimination metrics ranging between 0 and 1, which increase when suitability scores are high on presence points and low on absence points. The Boyce index is a reliability metric ranging between -1 and 1, only sensitive to suitability scores at presence points. The decision on model exclusion is based on the Boyce index, as recommended in presence-pseudo absence models (Leroy et al., 2018). For ensemble models, models for which predictions were worse than random were excluded (Breiner et al., 2015; Scherrer, Christe & Guisan, 2019), i.e. when the Boyce index was below 0.

Clamping masks showing the areas where climates are novel in the future (similar to a Multivariate Environmental Similartiy Surfaces analysis; Elith, Kearney & Phillips, 2010) were provided to determine whether models are well informed for predictions on future data.

To identify priority areas for conservation while accounting for uncertainty, a map was provided for each period (dry and wet seasons, present and future) showing the areas that are the most consistently identified as suitable between model replicates and scenarios Following Kujala et al. (2013), the weighted mean predictions discounted with inter-model variability (standard deviation) was computed for present and future predictions separately.

*Habitat and land use data*

Non-climatic habitat requirements of the modelled species were accounted for by applying a filter to the projected climate suitability based on land use and land cover data (Gillard et al., 2017). This enabled habitat to be accounted for without increasing model complexity while remaining biologically realistic and relevant for conservation applications. One challenge in modelling lateral-migrant Amazon fish species is that their distribution is highly variable within the year, as a result of important differences in hydrological regimes and behavioural adaptations. During the high-water period, *Arapaima* spp. colonize the main river channel and the flooded forests before returning to floodplain lakes to reproduce during the low-water period. Floodplain lakes thus provide key habitats for *Arapaima* spp. and must be prioritized for conservation (Richard et al., 2018). By contrast, the habitat types of the main channel (e.g. water colour types reflecting geological conditions; Junk et al., 2015) are also important determinants of fish diversity owing to differences in nutrient content (Oberdorff et al., 2019). According to the database used, *Arapaima* spp. seem almost exclusively found in white waters, which are nutrient rich, with high rates of biological production, sustaining large prey fish populations and favourable for omnivore-piscivorous fishes such as *Arapaima* spp. Note that *Arapaima* spp. can be occasionally found in black waters, but in very low densities (Hallwass, Schiavetti & Silvano, 2020). All terrestrial lands were removed and a set of projections for the flooded and the dry periods were produced. High-resolution data (3 arcsec) on wetland areas for the flooded and the dry season from were retrieved from Hess et al. (2015). Model predictions were resampled at the resolution of the wetland data to prevent any information loss. Data on water type (Venticinque et al., 2016) were also used to remove the black and clear waters from the predictions, to retain only white waters. As data on water colour types are linear (vector, i.e. no superficial data), a 0.5° buffer around lines was produced and model predictions falling outside the buffer were filtered.

Spatial features (barriers) that would prevent potential distribution shifts were considered. Typological data on dams obtained from Anderson et al. (2018) and the Brazilian National Water Agency (ANA; Accessed on December 2020; available at: http://www2.ana.gov.br), and waterfall locations from Oberdorff et al. (2019). These features were projected on predictions maps to examine their potential impact.

*Conservation gap analysis*

The extent of protected and unprotected suitable environments for the species were assessed for both flooded and dry periods, and for present and future conditions (2070). Information on the protection status was obtained from the World Database on Protected

Areas (UNEP-WCMC, 2019). Protection gap analyses usually involve binary transformations to be able to quantify the area potentially occupied by the species (e.g., De Carvalho et al., 2017; Bosso et al., 2018; Ahmadi et al., 2020; Hoveka, van der Bank & Davies, 2020). However, binary transformations are not recommended, as they are strong drivers of uncertainty (Muscatello, Elith & Kujala, 2020). Therefore, the frequency distribution of suitability scores inside and outside PAs is shown (Mod et al., 2020). As an indicator of overall suitability, the proportion of total environmental suitability inside and outside PAs is also provided.

## Results

*Present distribution modelling*

Model reliability varied with the variables included in small model modalities (Figure S2), with a median Boyce index of 0.30 before, and 0.47 after sample bias correction; 498 poorly performing models (Boyce index <0) in the uncorrected group and 408 in the corrected group (out of 1,500 per group) were discarded. The effect of sample bias correction was moderate, slightly higher than the variability between model replicates (relative overlap index = 0.06). Four uncorrelated variables were selected, all showing potentially high importance (Figure S3). The selected variables were, both for uncorrected and corrected groups, annual mean temperature (bio1), isothermality (bio3), temperature seasonality (bio4) and temperature annual range (bio7; Figure S4). The species is found in the warmest places of the Amazon basin, where annual mean temperature >21°C with an optimum near 28°C (i.e. the maximum annual mean temperature of the study area; Figure S4). Occupied areas are also characterized by high diurnal variation in temperature, low seasonality and intermediate annual variation in temperature relative to the Amazon basin (temperature annual range between 6 and 12°C; Figures S4, S5). According to the models, the most suitable conditions are met in the central and most eastern parts of the basin along the Amazon mainstem and the downstream part of its major tributaries, in the south-western part of the basin (i.e. Ucayali and Juruà rivers and their tributaries) and in its southern part along the Madre de Dios, Guapore and Madeira rivers.

*Future climate suitability*

Future predictions mostly depended on the emission scenario (RCP2.6 versus RCP8.5; Figure 1). In all cases, the models predict an important decline in climate suitability in most parts of the present range of the species and suggest, for RCP2.6, a shift in the suitable climate conditions towards the southern part of the basin (mostly tributaries located south of the Amazon mainstem). The decline in climate suitability is much stronger under RCP8.5 where suitability is projected to be optimal only in the most north-western part of the basin. Despite the apparent variability between models, predictions penalized by uncertainty overall agreed with unpenalized predictions (Figure 2). The clamping masks show an important area for which values were extrapolated for RCP8.5, mostly driven by bio1 (Figure S6). After removing all non-wetland and keeping only white waters, the total environmental suitability will decrease by 16.6% on average by 2070 during the high-water stage and by 19.4% during the low-water stage (Figures. 3; S7).

Most dams and dam projects are localized towards the edges of the Amazon basin, near the Andes. One dam is located west of the Amazon, potentially preventing movements between the central Amazon and the Peruvian Amazon (i.e. between the Amazon mainstem and the Ucayali and Marañòn rivers). Population movements will be mostly hampered in the Madeira River, where five dams or dam projects and waterfalls are located. These may represent dispersal barriers between the central and Bolivian Amazon (Figures S7, S8).

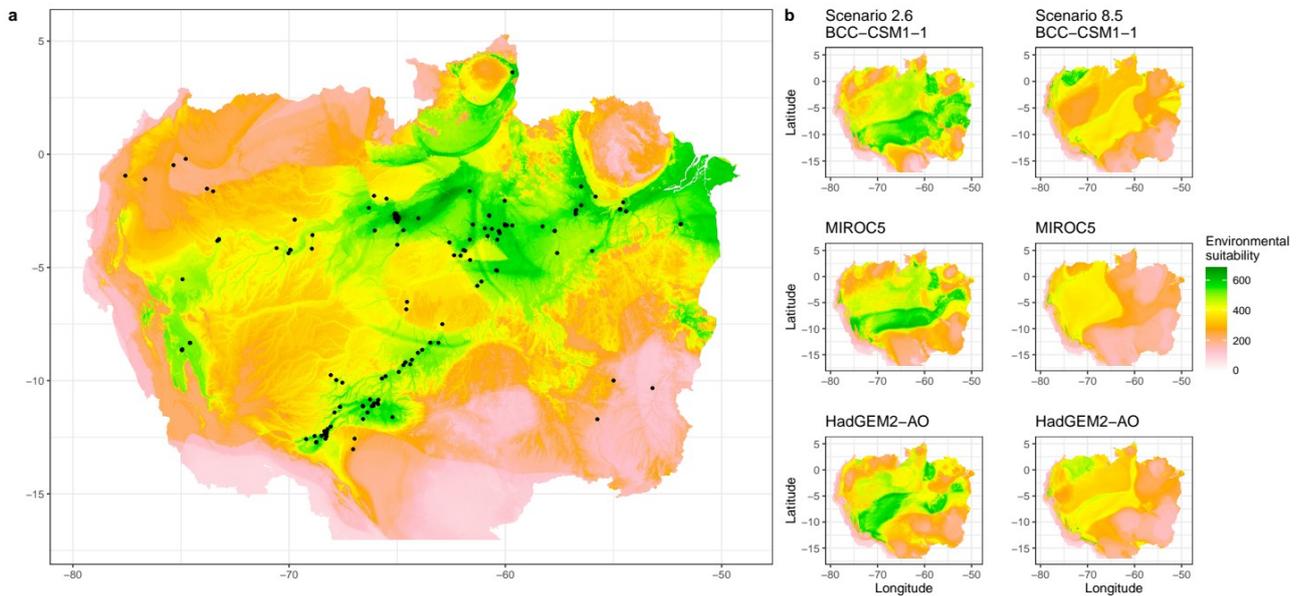

**Figure 1. Present (a) and future 2070 (b) climate suitability for *Arapaima* spp. in the Amazon basin. Suitability scores were estimated from ensemble of small models for two emission scenarios and three global circulation models, and corrected for sample bias. Black points represent occurrence records. Axes represent the coordinates (WGS84).**

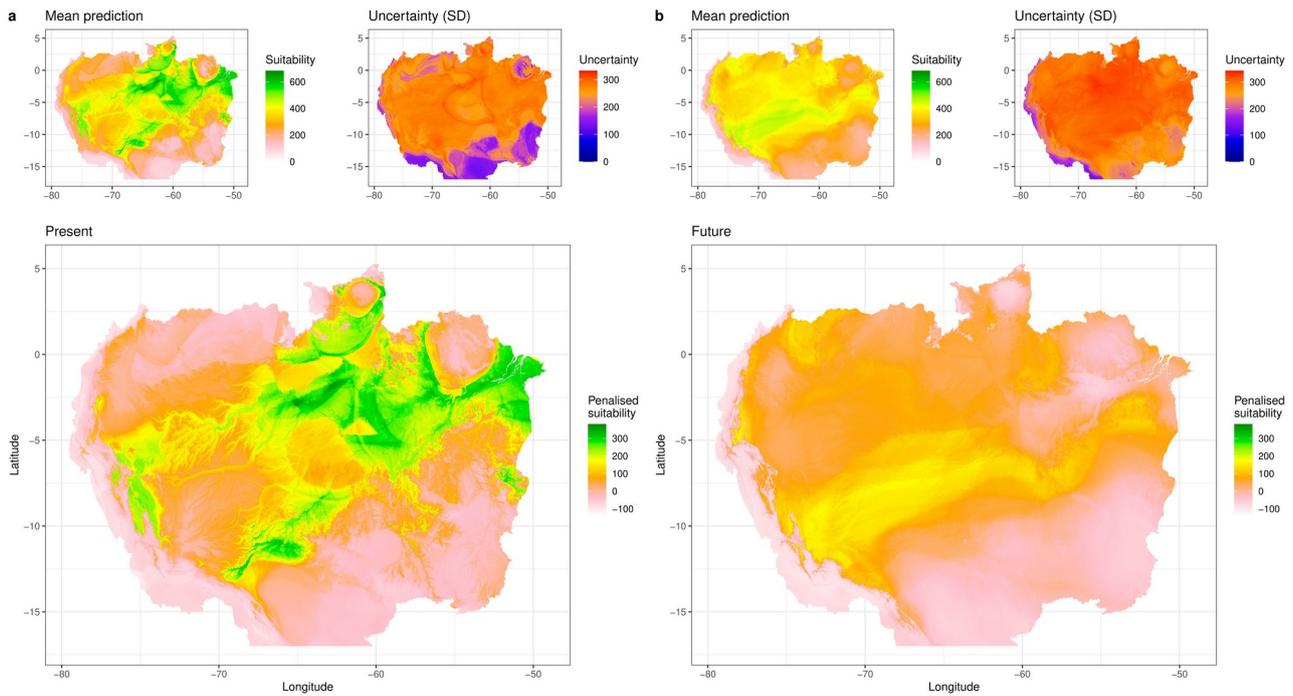

**Figure 2. Present (a) and future 2070 (b) predicted climate suitability for *Arapaima* spp. accounting for uncertainty. Bottom panels are the result of mean predictions discounted with inter-model variability (standard deviation, as an indicator of uncertainty related to model settings and climate data).**

*Gap analysis*

Under present conditions, 31.4% of the total suitable environment is included within PAs during the high-water stage, and 30.7% during the low-water stage. In the future, the proportion of protected suitable areas will decrease by approximately 5%, with 26.3% and 25.9% of suitable environments within PAs for the high- and low- water stages, respectively (Figures 3, S9–S11). The frequency distribution of suitability scores shows that the majority of the most suitable areas (score >300) remains unprotected (see grey area in Figure 3). The most important gap in regard to future environmental suitability is located between the upper parts of the Ucayali, Juruà and Purus rivers (south west to the mainstem).

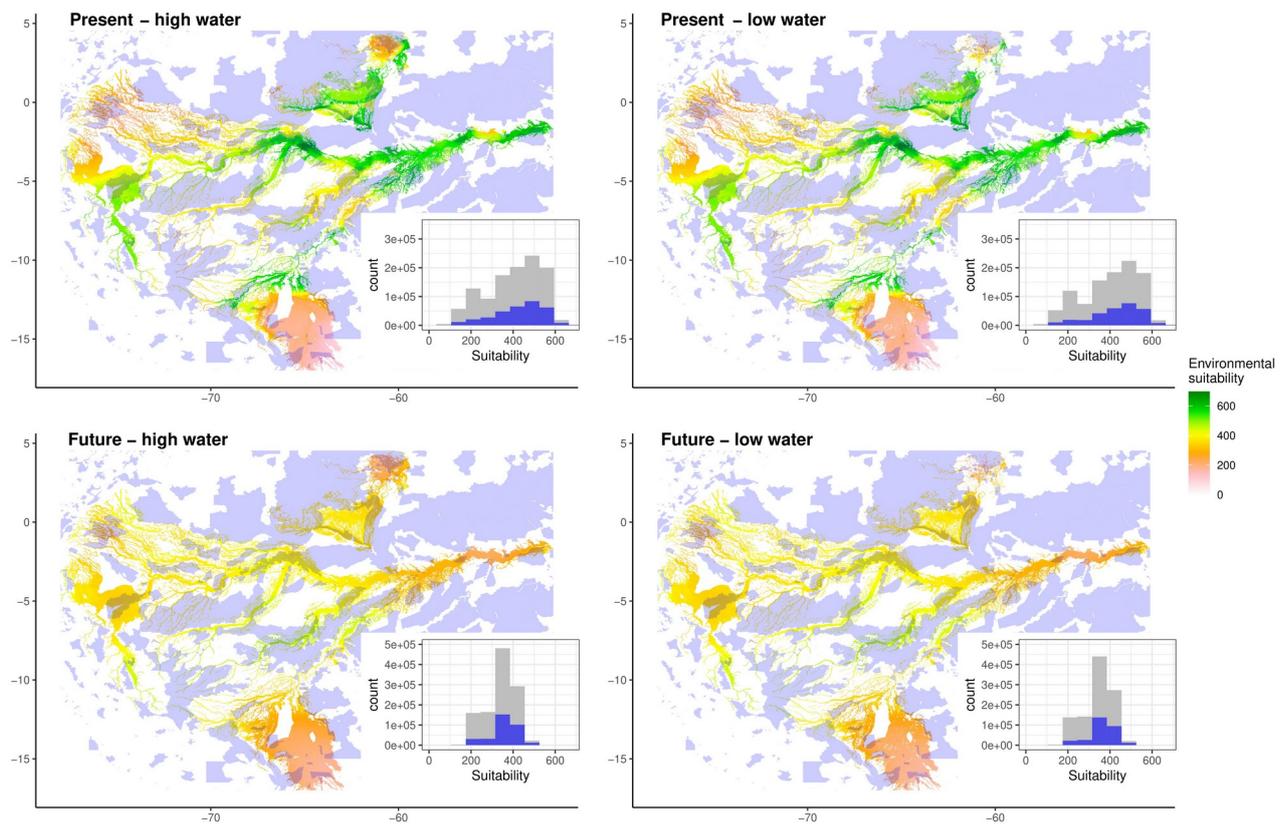

**Figure 3.** Predicted environmental suitability for *Arapaima* spp. for present (top) and future 2070 (bottom) conditions at two stages of water levels, with protected areas (blue). The frequency distribution of environmental suitability scores inside (blue) and outside (grey) protected areas is shown.

## Discussion

A robust approach for species with scarce distribution data was used (ensemble of small models), based on climate as predictors and habitat as filters, accounting for sample bias, hydroperiod and dispersal barriers, and integrating uncertainty by discounting inter-model variability from predictions, and the distribution of *Arapaima* spp. was modelled successfully. The species complex may face important environmental degradation driven by climate change by 2070. Significant conservation gaps in the coverage of suitable environments were identified, which are likely to worsen with climate change.

*Drivers of Arapaima spp. distribution*

*Arapaima* spp. are found in the hottest lowland areas of the Amazon basin, with high daily variation and intermediate variability in annual temperatures overall. Our predictions fit well the known distribution of the species complex, apart from the populations located in the extreme north west of the Amazon basin (corresponding to the Ecuadorian and Colombian Amazon), where climate conditions were shown to be sub-optimal. Overall, the models based on temperature identified well the riverine areas (Figure 1), a sign of good model performance. The minimum annual mean temperature where the species can be found (21°C) is consistent with the findings of Lawson et al., (2015) who found that *Arapaima gigas* ceased feeding at an average temperature of 20.8°C. The high daily variation may be related to heavy rain episodes, which is consistent with thedependence of Arapaima spp. on floodplains. The low seasonality corresponds to equatorial regions where the species will not be exposed to low temperatures (detrimental to *Arapaima* spp.) during any period of the year.

Suitable conditions were found throughout the Amazon River, but *Arapaima* species are highly restricted to specific habitats (deep and large slow-flowing waters that are connected to lateral water bodies; Arantes et al., 2013). Those variations could not be perceived at the resolution of the environmental variables used (3 arcsec, approximately 100 m). Therefore, our predictions must be regarded as regions where the climate is suitable, but local habitat requirements must be considered in future sampling campaigns (e.g. Arapaima species are restricted to deep waters). Another limitation of the models is that they ignore whether Arapaima spp. will tolerate the upcoming novel conditions. Future research should assess the species' capacity for acclimatization, plasticity and adaptation to changing conditions. Precipitation variables were not included in the models because we believe that they are drivers of uncertainty (Kent, Chadwick & Rowell, 2015; Kim et al., 2020; Marra et al., 2021). Nevertheless, precipitation is usually important in river systems as it may affect the discharge regime and consequently the degree of seasonal connectivity between the river and the floodplain. Topographic variables that may constrain Arapaima spp. habitat suitability (e.g. elevation, river slope and Strahler's order; Kuczynski, Legendre & Grenouillet, 2018) were not included; however, these were indirectly taken into account in the models as being strongly linked with the presence/absence of floodplain areas in the basin.

*The future of* Arapaima *spp.*

A geographical shift in suitable climatic conditions by 2070 in the most optimistic scenario, and a generalized decline in climate suitability in the worst-case scenario is predicted. In the optimistic scenario (RCP-SSP 2.6), suitable conditions will be met towards the upper part of the Ucayali, Juruà and Purus rivers and their tributaries, where large floodplains are located. In the most pessimistic scenario (RCP-SSP 8.5), the same region was identified as the most suitable, but conditions will be largely sub-optimal. Models also predict a species distribution shift towards tributaries located south of the Amazon mainstem and a northward shift of south-western populations (mostly in the Bolivian part of the basin). These potential shifts may be hampered by the presence of two dams between the upper and the lower Madeira River. However, in this specific case, Arapaima spp. were not originally present in the upper Madeira and are at present considered invasive in this region so that the presence of these barriers should not be viewed as a strong conservation problem. The results of this study differ overall from those obtained by Oliveira et al. (2020) who found that climate conditions may remain suitable in the western part of the Amazon basin. It should be noted that our predictions are nevertheless in agreement with the potential decline in suitable conditions in the central Amazon, and for the shift in suitable conditions at the extreme south-western part of the Amazon basin (but remaining sub-optimal in our case). Differences may have been partly driven by the occurrence data (all the recently available data from the AmazonFish database were used). The input climate data (baseline/data source and GCMs) used can be a strong source of uncertainty as well (Baker et al., 2016; Dubos et al., 2022a). In their study, Oliveira et al. (2020) used Worldclim data with the CCSM4 GCM while CHELSA with three different GCMs were used here, showing that the choice of GCM may affect predictions. The use of multiple GCMs was highly recommended as it represents an important source of uncertainty in model projections (Buisson et al., 2010; Dubos et al., 2022a). This also applies to the present study, with notable differences throughout the Amazon basin (Figures 1b, 2b). The present results also differ for the south-western part of the Amazon Basin. This region was discarded by the absence of white waters, which is a key habitat feature for *Arapaima* spp., compared with Oliveira et al. (2020) who projected a high suitability in the future.

The certainty of our future predictions was limited by novel climate conditions (mostly driven by bio1) for which models were uninformed, inducing extrapolations. Given the shape of the response curve to bio1 (higher suitability in the hottest conditions; Figure S4), extrapolated predictions may result in an overestimation of climate suitability when bio1 is above present conditions. Hence, in the case where future conditions are beyond the thermal tolerance of *Arapaima* spp., the decline in climate conditions may be even worse than expected here. Further studies should assess the upper limit of thermal tolerance for *Arapaima* spp.

*Conservation considerations*

The Amazon region is affected by poverty and limited food security, with more than 30 million people depending on floodplain fish for food (Petersen et al., 2016). After decades of industrial harvesting pressure, the Arapaima species complex was considered overexploited and declining as a result of overfishing (Isaac, Rocha & Mota, 1993; Castello, Stewart & Arantes, 2011). The decline in mean catches has already been reported in Brazil (Castello, Stewart & Arantes, 2011) and Peru (García Vásquez et al., 2009). Conservation measures were then undertaken in various regions, consisting of banning industrial fishing activities in selected oxbow lakes (Castello et al., 2013). These measures proved effective (Petersen et al., 2016) but the fishing pressure persisted (Castello et al., 2015; Cavole, Arantes & Castello, 2015). Recent work has shown that the involvement of local communities enabled promising outcomes to be reached regarding the management of Arapaima spp. (Gamarra et al., 2022). We stress the need to develop collaboration between scientists, local stakeholders and indigenous communities to ensure the sustainability of resources and allow traditional fishing activities to persist.

Climate change will bring an additional extinction risk to the species complex and probably most floodplain-dwelling species, which may be further worsened with future dam building (Winemiller et al., 2016). This might be the case for most large fish species in the Amazon River, including species that are frequently caught by fishers. The majority of suitable environments of Arapaima spp. remains currently unprotected and the proportion of unprotected suitable areas is, as shown here, likely to increase with climate change. This may apply to the wider freshwater fish community inhabiting floodplains (Frederico et al., 2021). The present PA network downplays the importance of freshwater species as well as important terrestrial biodiversity during the terrestrial phase corresponding to the low-water stage (Piedade et al., 2010). Arapaima is threatened by deforestation by the alteration of carbon cycles and the decline of food resources (Carvalho et al., 2018), which highlights the need to integrate buffer zones of forested areas in the floodplain surroundings within the PA network. The most important gap with regard to future environmental suitability is located between the upper parts of the Ucayali, Juruà and Purus rivers. We thus recommend the extension of the existing PAs (e.g. Vale do Javi southward, Kanamari do Rio Juruà, El Sira and the PA network surrounding the Reserva Extrativista Do Médio Purus) to the floodplains surrounding those rivers and their tributaries. However, merely expanding the PA system is not sufficient as the capacity of this network to protect freshwater biodiversity remains unclear. Indeed, human pressure on many PAs has increased, suggesting a gap in their management with regard to halting habitat loss and intensified human use (Azevedo-Santos et al., 2019; Duponchelle et al., 2021). In addition to the expansion of the PA network, we therefore encourage the strengthening of community- based local management (Gamarra et al., 2022) initiatives. This will help to promote the importance of floodplain species for their intrinsic conservation value, for their economic importance and as food resources, and raise awareness regarding climate change effects.

## ACKNOWLEDGEMENTS

This study was funded through the 2017-2018 Belmont Forum and BiodivERsA joint call for research proposals, under the BiodivScen ERA-Net COFUND programme, and with the funding organisations French National Research Agency (ANR), São Paulo Research Foundation (FAPESP), National Science Foundation (NSF), the Research Council of Norway and the German Federal Ministry of Education and Research (BMBF). We thank Boris Leroy for useful discussions and Lucie Kuczynski for her valuable comments on the manuscript.

## DATA AVAILABILITY STATEMENT

Occurrence data are available at:

https://figshare.com/articles/dataset/A_database_of_freshwater_fish_species_of_the_Amazon_Basin/9923762

## AUTHOR CONTRIBUTION

**Nicolas Dubos**: Conceptualisation (equal), data curation (lead), formal analysis (lead), methodology (lead), visualisation (lead), writing – original draft preparation (lead), writing – review & editing (equal). **Maxime Lenormand**: Conceptualisation (equal), funding acquisition (equal), project administration (equal), supervision (equal), methodology (equal), visualisation (equal), writing – review & editing (equal). **Leandro Castello**: Conceptualisation (equal), writing – review & editing (equal). **Thierry Oberdorff**: Conceptualisation (equal), writing – review & editing (equal). **Antoine Guisan**: Conceptualisation (equal), methodology (equal), writing – review & editing (equal). **Sandra Luque**: Conceptualisation (equal), funding acquisition (equal), project administration (equal), supervision (equal), writing – review & editing (equal).

# Supporting information

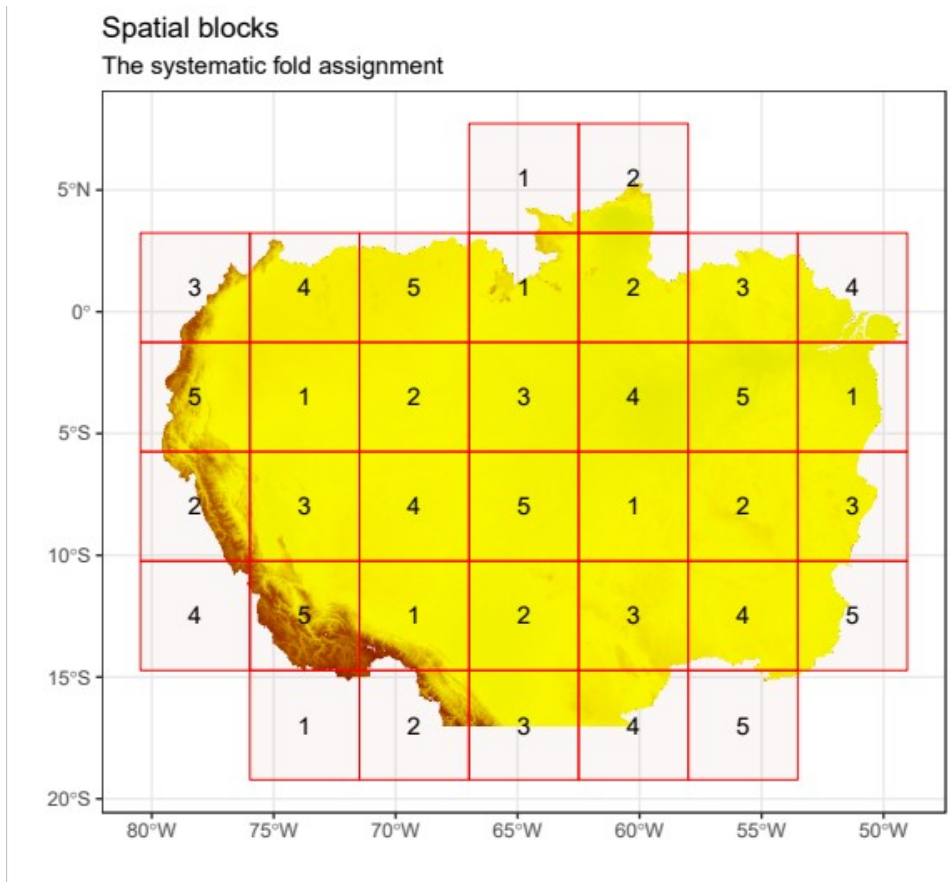

**Figure S1. Block-cross validation folds for the modelisation of Arapaima sp.**

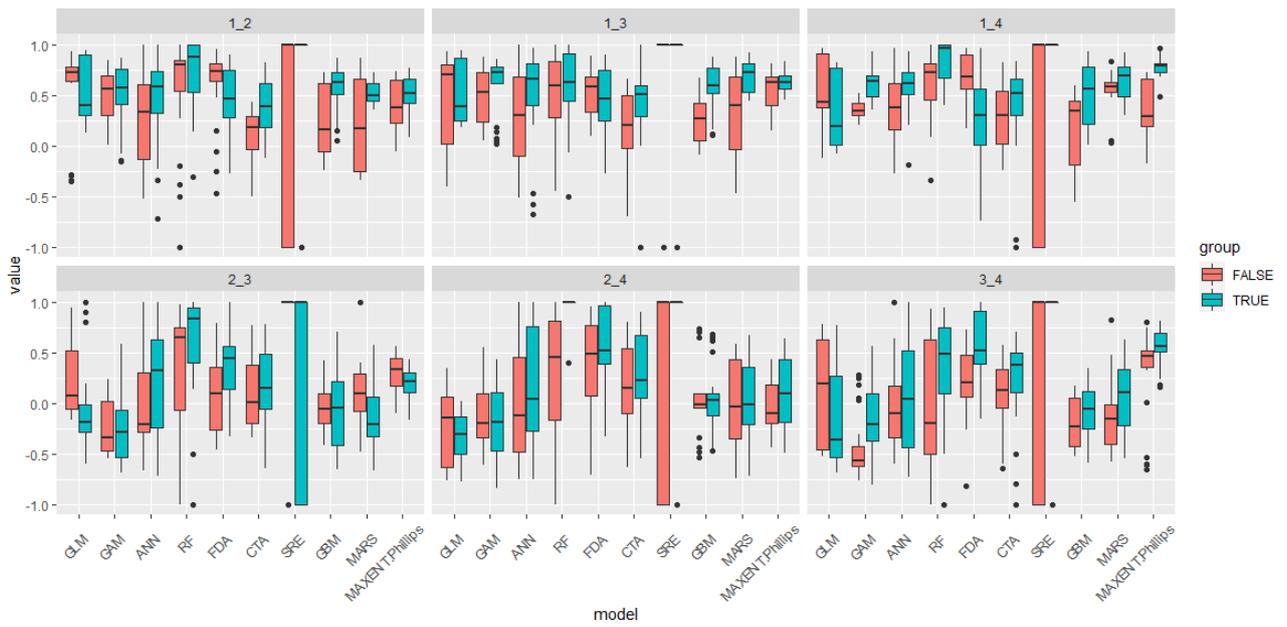

**Figure S2.** Model reliability (Boyce indices) for each ESM, with (blue) and without (red) sampling bias correction.

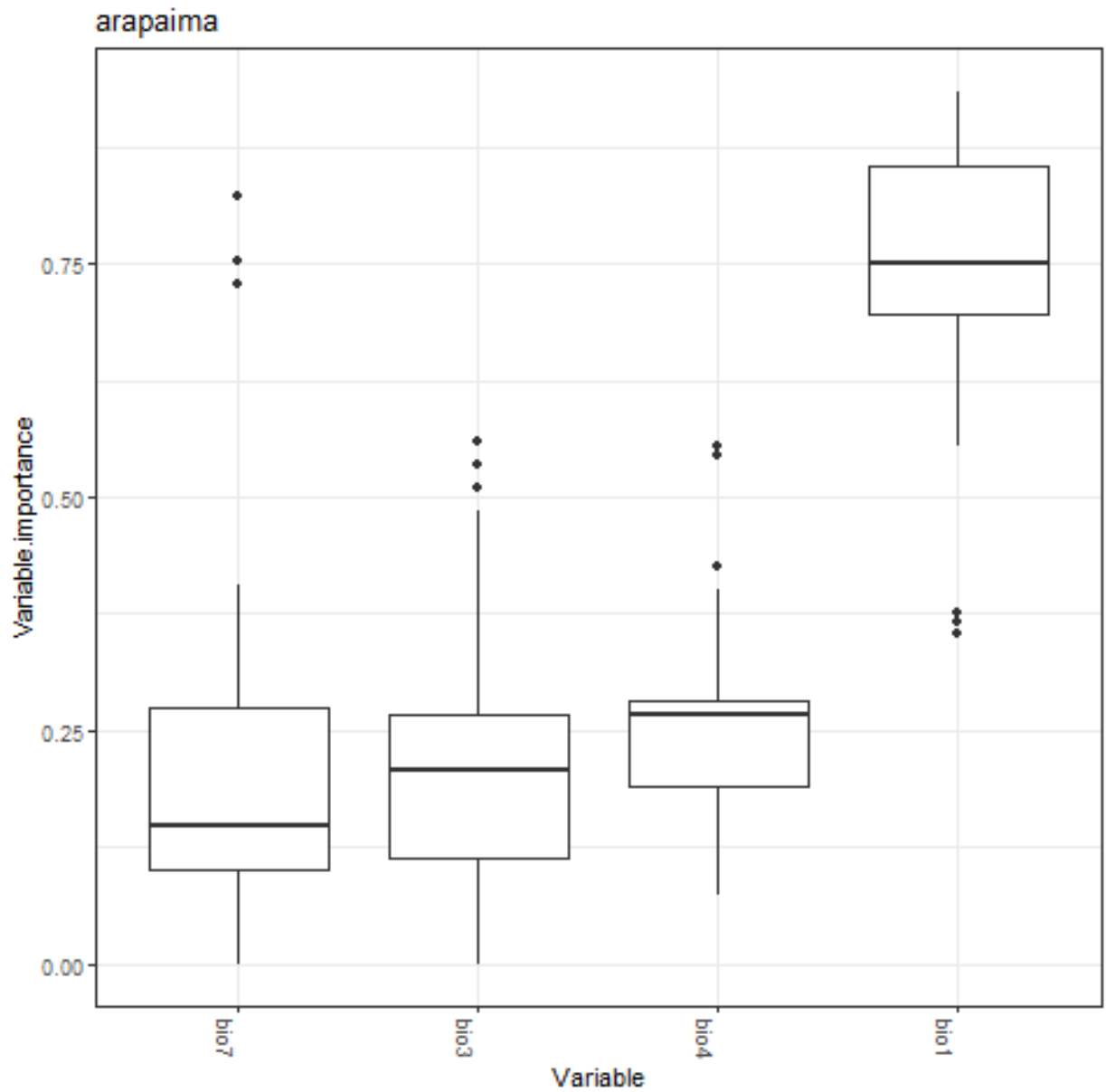

**Figure S3. Variable importance for the species distribution model of Arapaima sp.**

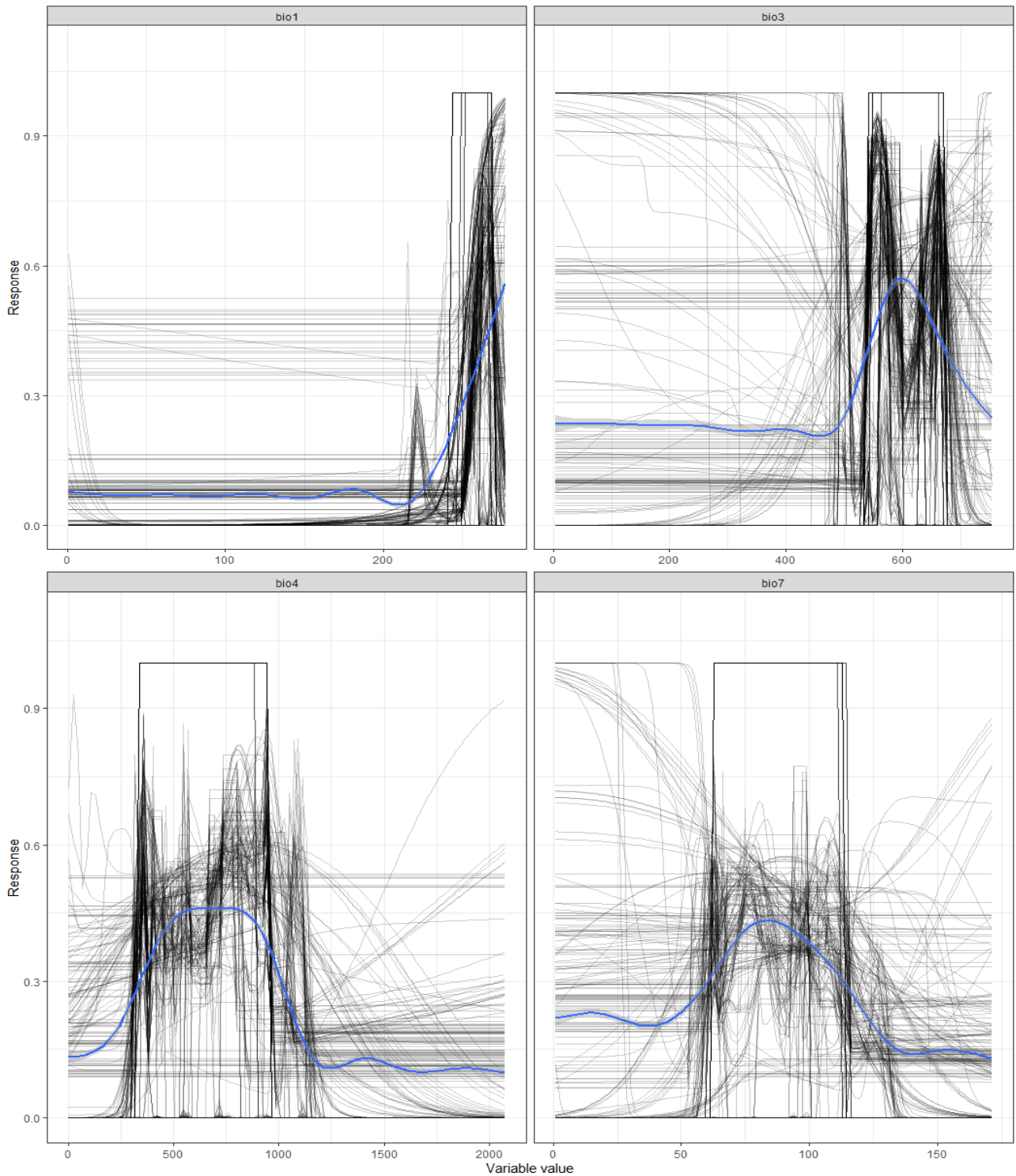

**Figure S4. Response curve to bio1 (annual mean temperature), bio3 (isothermality), bio4 (temperature seasonality) and bio7 (temperature annual range) for uncorrected models. Black lines represent predicted values for each individual model. The blue line represents the smoothed response across all models.**

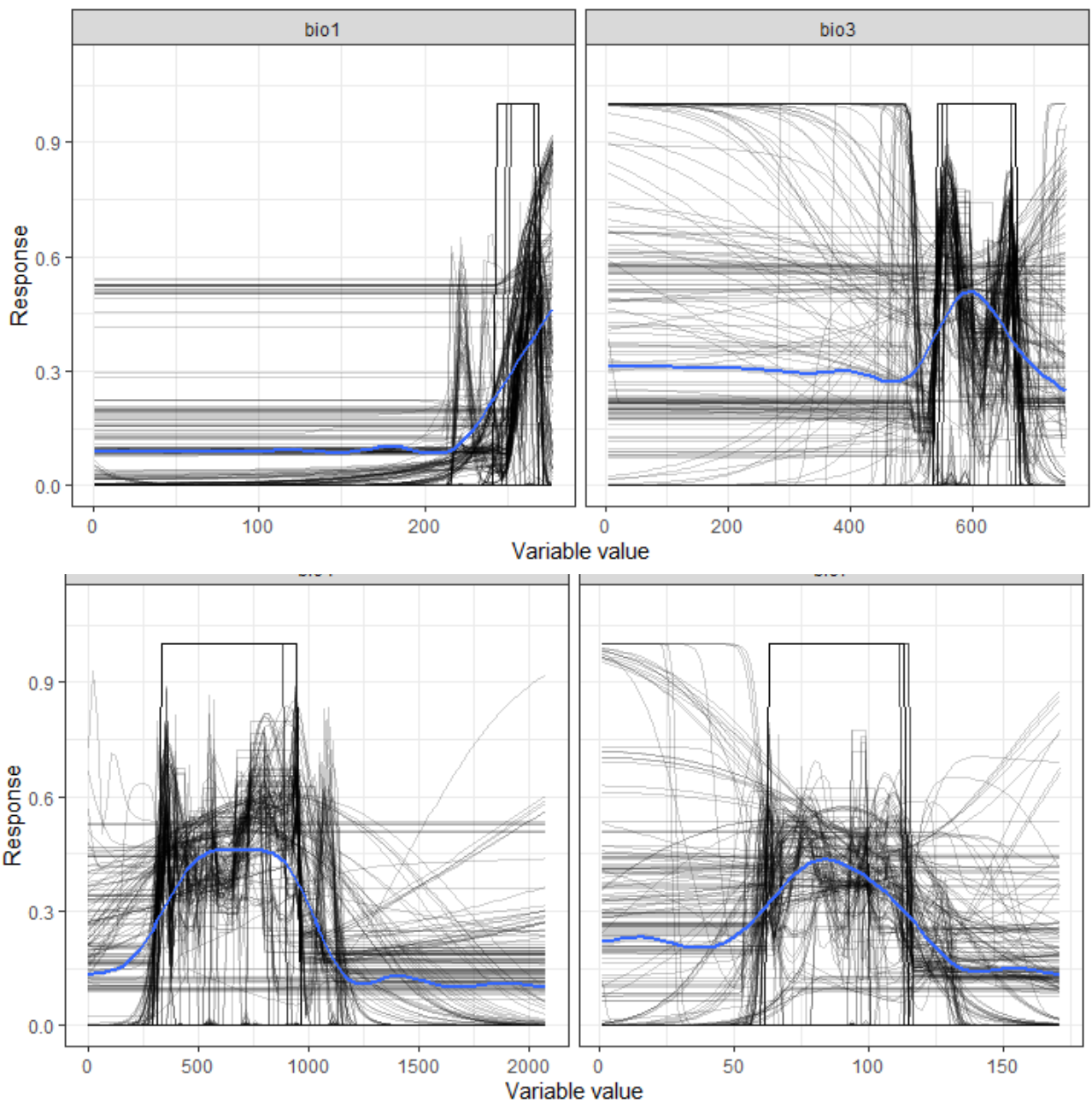

**Figure S5. Response curves to bio1 (annual mean temperature), bio3 (isothermality), bio4 (temperature seasonality) and bio7 (temperature annual range) for corrected models. Black lines represent predicted values for each individual model. The blue line represents the smoothed response across all models.**

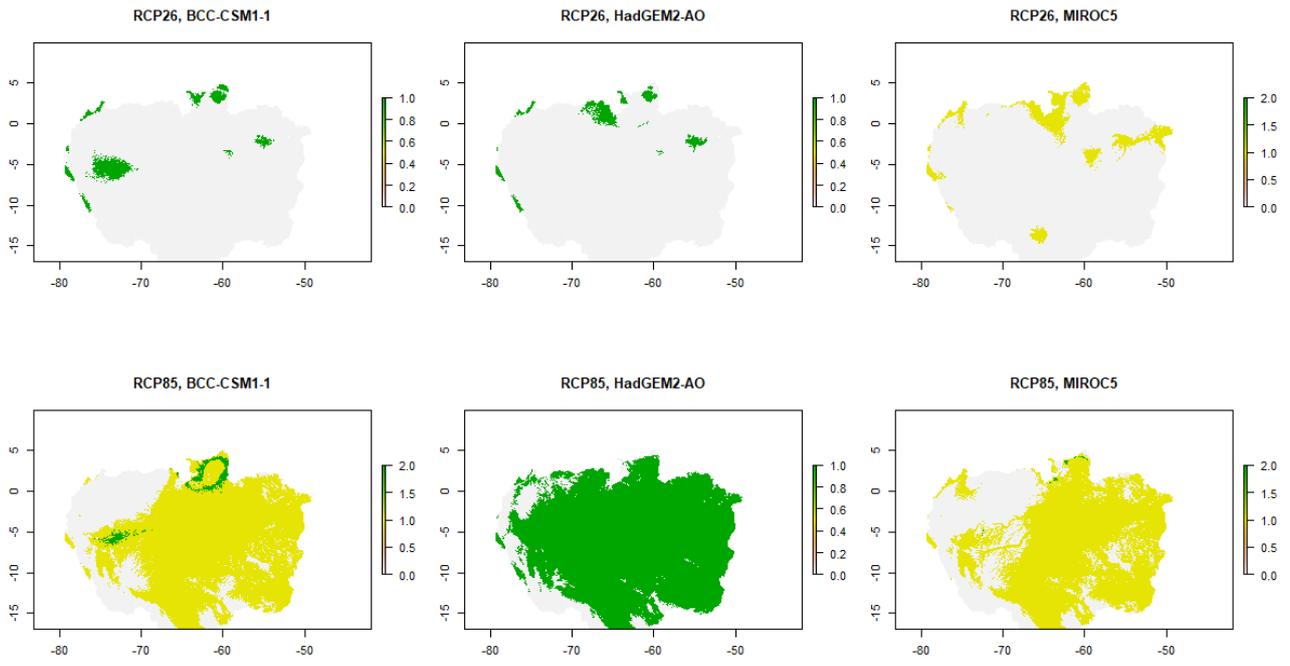

**Figure S6. Clamping mask for the distribution model of *Arapaima* sp. showing novel climate conditions for 2070. Novel conditions are mainly driven by temperature annual range (bio1).**

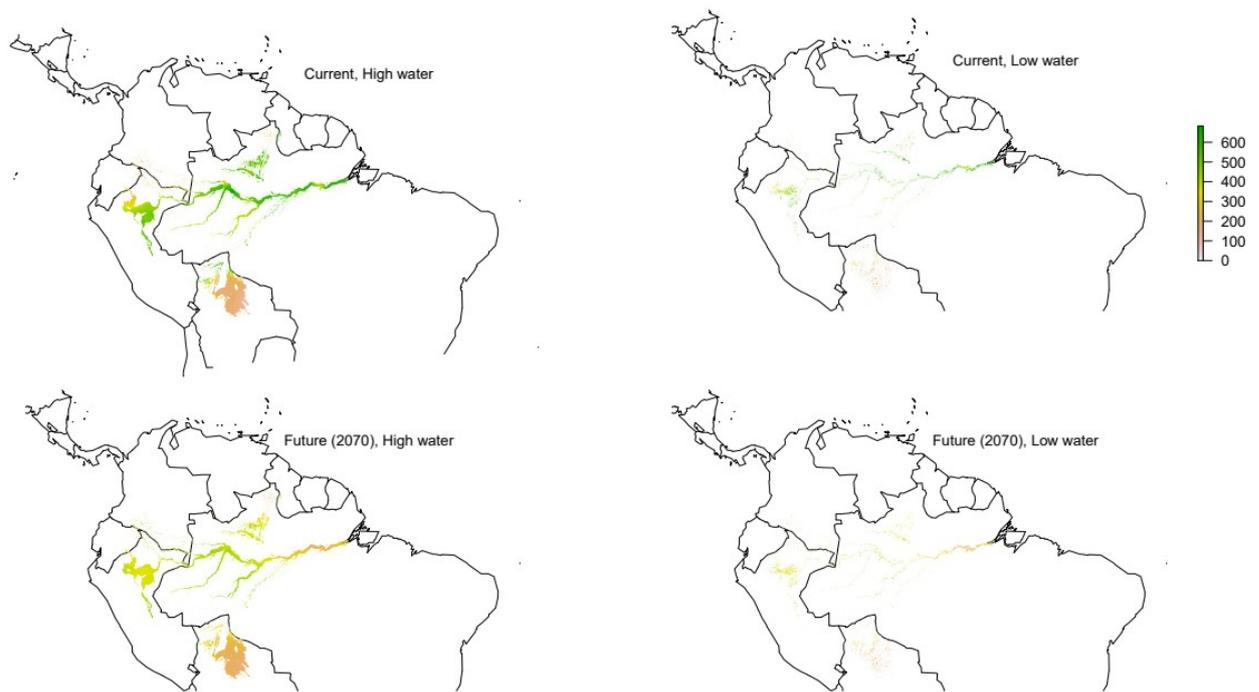

**Figure S7. Present and future environmental suitability for *Arapaima* spp. Projections accounted for hydrological period (high versus low water stage) and water colour (white waters only). Black lines represent the administrative borders.**

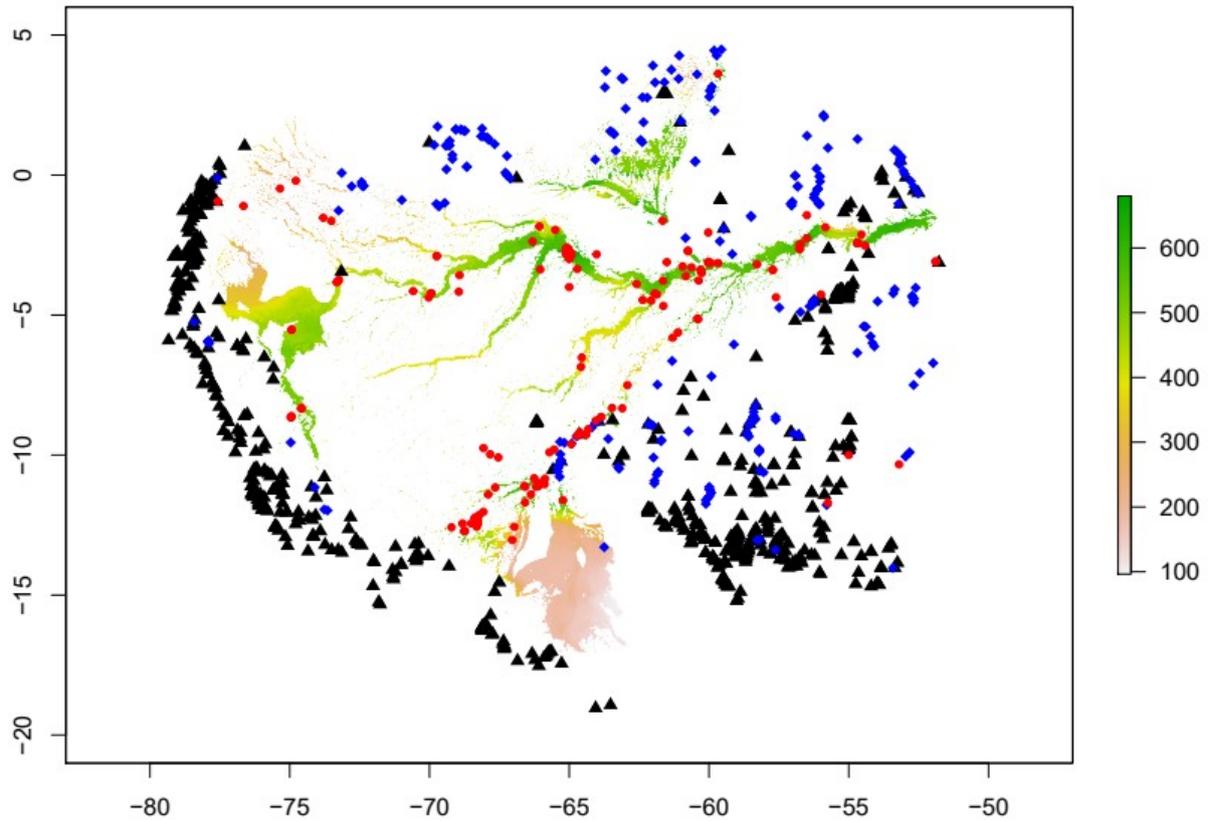

**Figure S8. Present environmental suitability for *Arapaima* spp. with dispersal barriers (blue diamonds: waterfalls; black triangles: dams). Red dots are the occurrence points. Axes represent geographic coordinates (WGS84).**

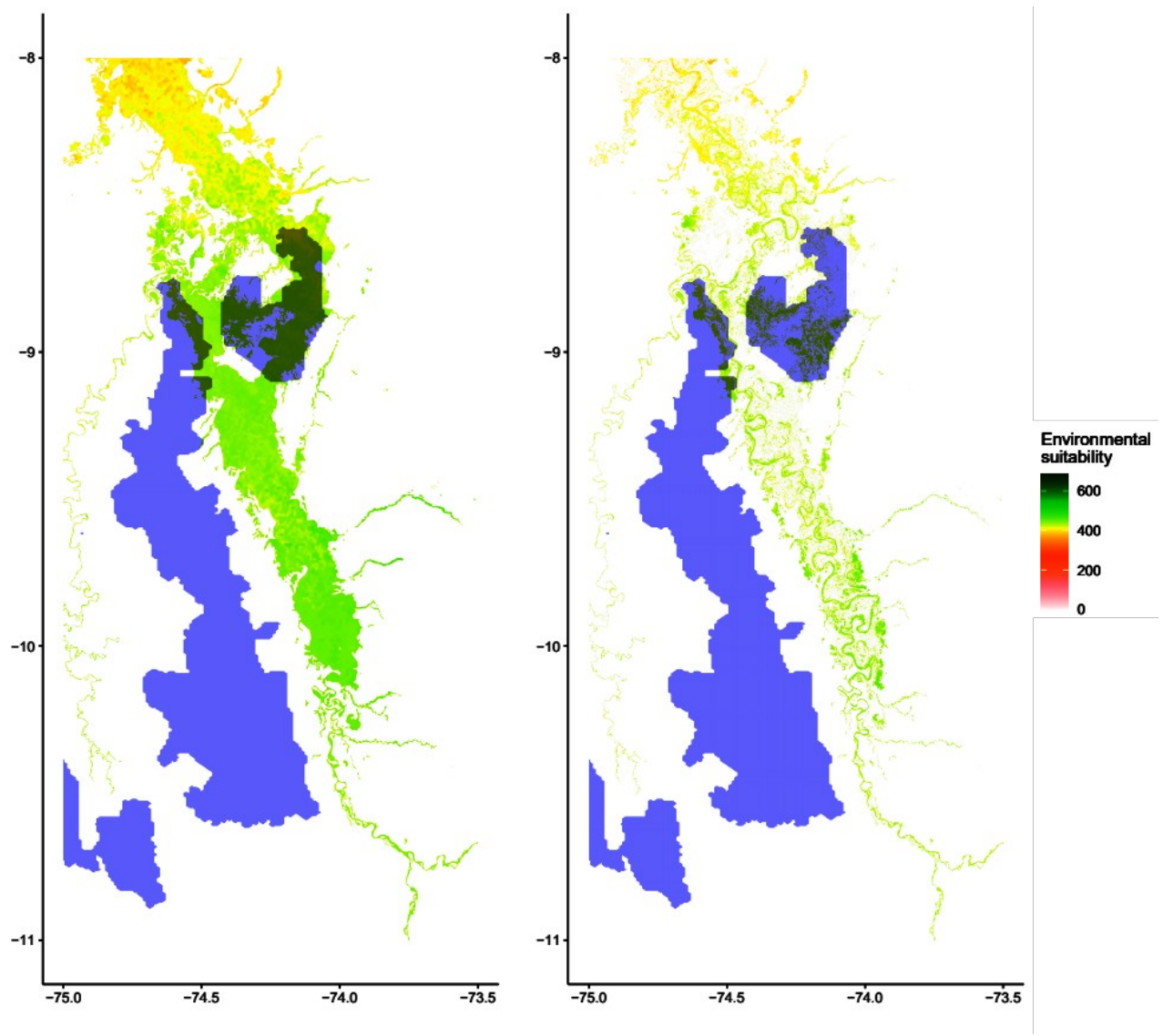

**Figure S9. Projected future environmental suitability for *Arapaima* sp. at two stages of water levels (left: high-stage; right: low-stage), with protected areas (blue) for the upper Ucayali river.**

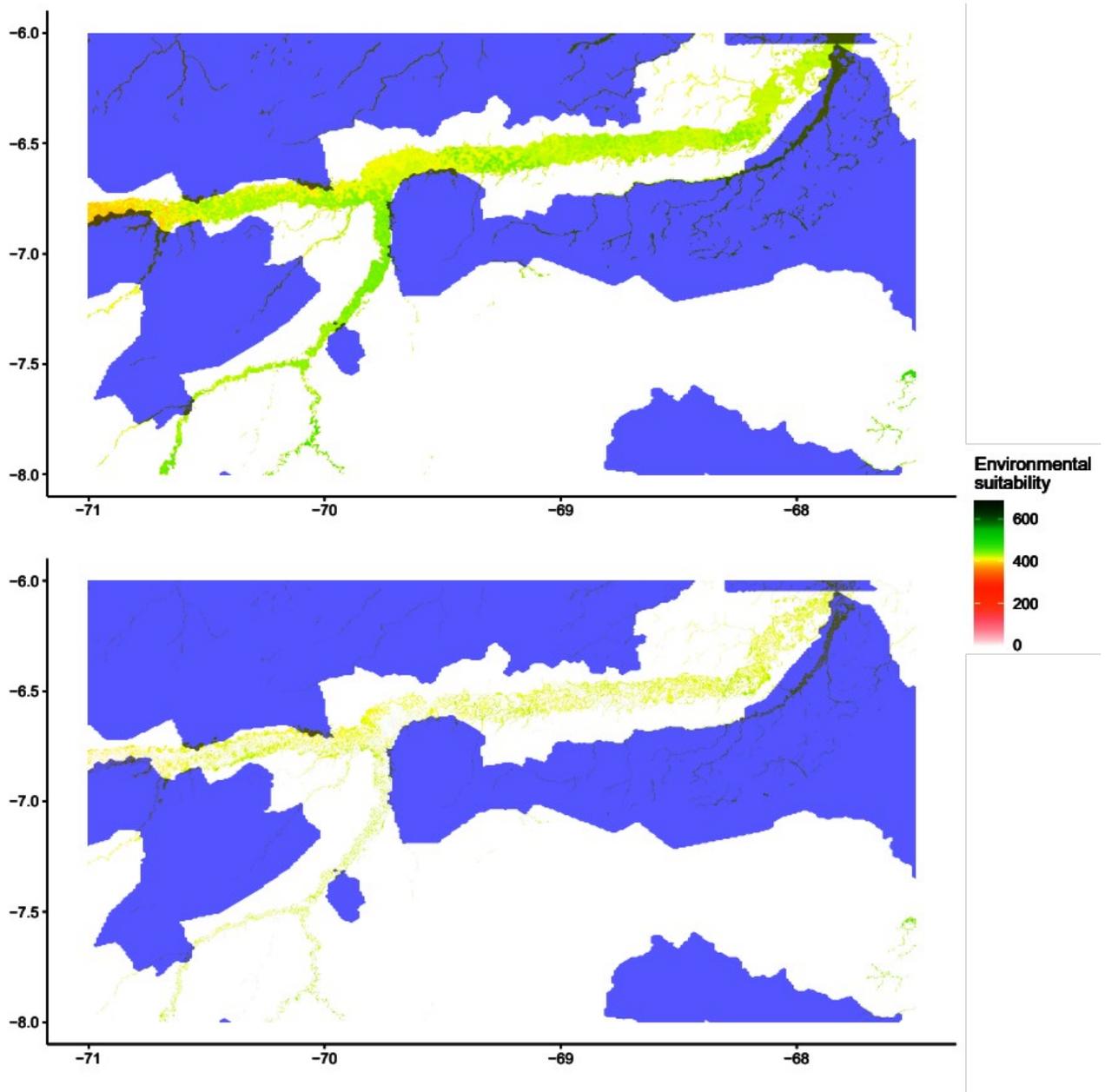

**Figure S10. Projected future environmental suitability for *Arapaima* sp. at two stages of water levels (top: high-stage; bottom: low-stage), with protected areas (blue) for the upper Juruà river.**

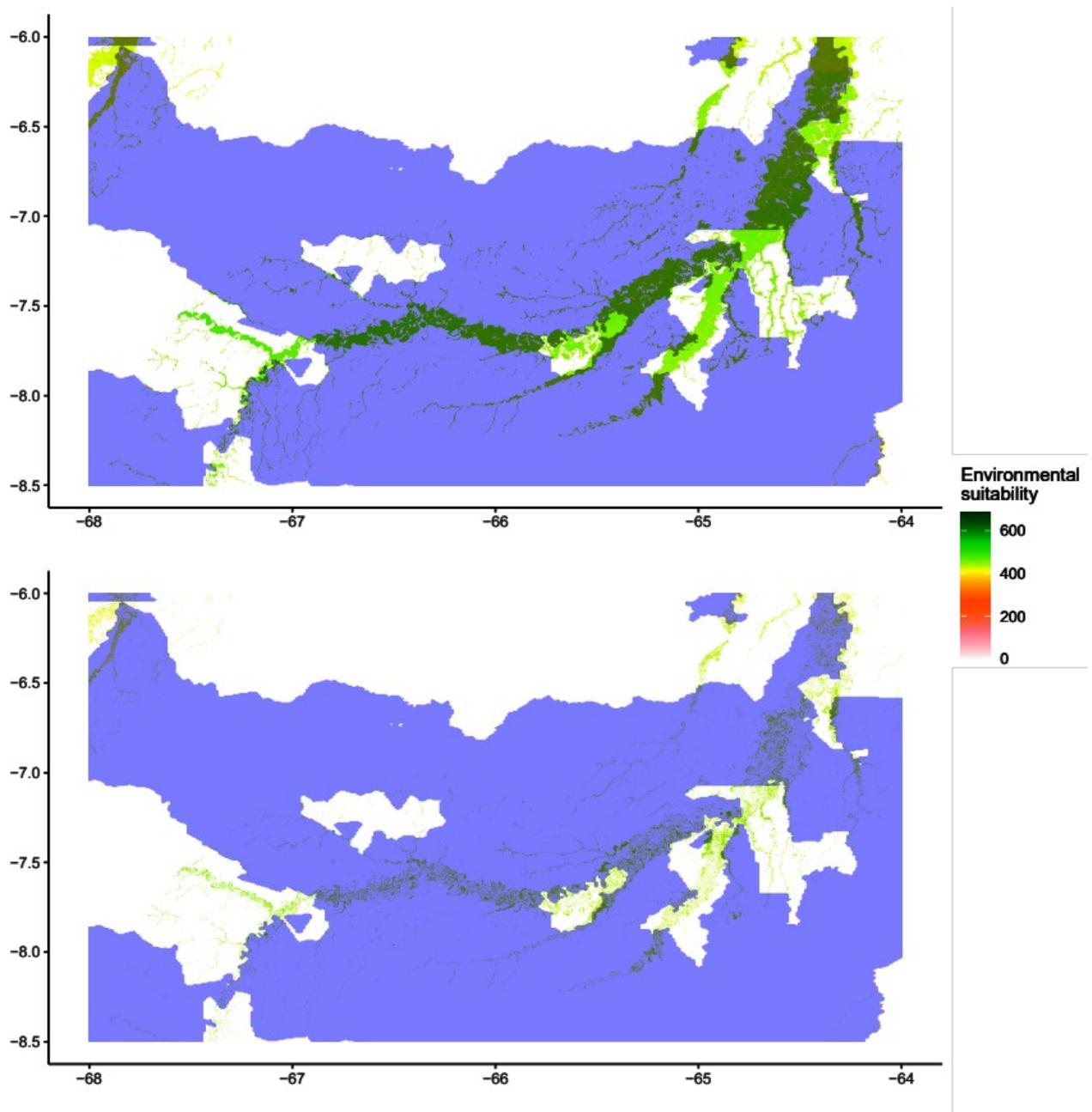

**Figure S11. Projected future environmental suitability for *Arapaima* sp. at two stages of water levels (top: high-stage; bottom: low-stage), with protected areas (blue) for the upper Purus river.**